\documentclass[prc,twocolumn,showpacs]{revtex4-1}
\usepackage{bm}
\usepackage{soul}
\usepackage{array}
\usepackage{lipsum}
\usepackage{relsize}
\usepackage{amssymb}
\usepackage{amsmath}
\usepackage{amsfonts}
\usepackage{mathrsfs}
\usepackage{graphicx}
\usepackage[usenames]{color}
\usepackage[mathscr]{euscript}

\newcommand{\ChiSq}[1]{\mbox{\large$\chi$}^{2}_{\mbox{\scriptsize$\raisebox{-2pt}{$\!#1$}$}}}
\newcommand{\On}[2]{\mbox{${\mathcal O}$}_{#1}^{\mbox{\scriptsize${\rm #2}$}}}

\begin{document}

\title{Analytic insights on the information content of new observables} 
\author{Wei-Chia Chen}\email{chenw@cshl.edu}
\affiliation{Simons Center for Quantitative Biology, 
Cold Spring Harbor Laboratory,  Cold Spring Harbor, 
NY 11724, USA}
\author{J. Piekarewicz}\email{jpiekarewicz@fsu.edu}
\affiliation{Department of Physics, Florida State University, 
               Tallahassee, FL 32306, USA}
\date{\today}
\begin{abstract}
Uncertainty quantification has emerged as a rapidly growing field in nuclear 
science. Theoretical predictions of physical observables often involve 
extrapolations to regions that are poorly constrained by laboratory experiments 
and astrophysical observations. Without properly quantified theoretical errors, 
such model predictions are of limited value. Also, one often deals with 
theoretical constructs that involve fundamental quantities that are not accessible 
to experiment or observation. Particularly relevant in this context is the pressure 
of pure neutron matter. In this contribution we develop an analytic framework to 
answer the question of ``How can new data reduce uncertainties of current theoretical 
models?"[P.-G. Reinhard and W. Nazarewicz, Phys. Rev. C81, 051303(R) (2010)].
Simple and insightful expressions are obtained to quantify the impact of one or two 
new observables on theoretical uncertainties in two critical quantities: the slope of 
the symmetry energy at saturation density and the pressure of pure neutron matter 
at twice nuclear matter saturation density.
\end{abstract}

\maketitle

\emph{Introduction.}
In the last few years we have witnessed historical discoveries in astronomy with far-reaching 
implications in nuclear astrophysics. First, the gravitational wave detection of the binary neutron 
star merger GW170817\,\cite{Abbott:PRL2017} together with  its associated electromagnetic 
counterpart\,\cite{Drout:2017ijr} are providing answers to some of the most fundamental questions 
animating nuclear science today\,\cite{LongRangePlan,QuarksCosmos:2003}. Second, pulsar-timing 
observations of the millisecond pulsar J0740+6620 by Cromartie and collaborators yielded the heaviest 
neutron star mass reported to date: $M\!=\!2.14^{+0.10}_{-0.09}\,M_{\odot}$\,\cite{Cromartie:2019kug}.
Finally, the first \emph{simultaneous} determination of  the mass and radius of a neutron star 
(J0030+0451) was reported by both the Maryland\,\cite{Miller:2019cac} and Amsterdam 
collaborations\,\cite{Riley:2019yda}. Using the Neutron star Interior Composition Explorer (NICER) 
aboard the international space station, pulse-profile modeling of the thermal emission from the 
pulsar's hot spots revealed a mass of about 1.4\,$M_{\odot}$ and a radius of nearly 13\,km, with 
a $\pm\,10\%$ uncertainty in both quantities.  

Whereas electromagnetic- and gravitational-wave detections are providing powerful constraints 
on the equation of state (EOS) at densities above nuclear matter saturation density, measurements of 
nuclear observables---such as neutron skins, electric dipole polarizabilities, and heavy-ion 
collisions---constrain the EOS near saturation density; see Refs.\,\cite{Tsang:2012se,Horowitz:2014bja,
Thiel:2019tkm} and references contained therein. It is this compelling connection between 
``heaven and earth" that promises unprecedented advances in the quest to determine the 
equation of state.

Besides these remarkable advances in nuclear astrophysics, uncertainty quantification 
has seen significant growth during the last decade. One of the earliest appeals to the 
theoretical community was issued by the Editors of the Physical Review A who stated: 
``it is all too often the case that the numerical results are presented without uncertainty 
estimates"\,\cite{PhysRevA.83.040001}. The nuclear theory community has responded 
to the challenge with a large number of publications that have addressed not only the 
critical role of theoretical uncertainties, but also the wealth of information that is contained 
in the study of statistical correlations among observables\,\cite{Kortelainen:2010hv,
Reinhard:2010wz,Fattoyev:2011ns,Fattoyev:2012rm,Piekarewicz:2012pp,
Reinhard:2012vw,Erler:2012qd,Fattoyev:2014pja,Dobaczewski:2014jga,Chen:2014sca,
Ireland:2015}. 

Our present work is motivated by the decade-old paper by Reinhard and 
Nazarewicz\,\cite{Reinhard:2010wz}. The paper poses two important questions on the 
uniqueness and usefulness of a new observable: (i) \emph{Considering the current theoretical 
knowledge, what novel information does new measurement bring in?} and (ii) \emph{How can 
new data reduce uncertainties of current theoretical models?} In an effort to illuminate this 
numerical procedure, we propose to answer these questions analytically and extend the 
argument to the case of adding more than one observable. Although by necessity some 
approximations will be required, our goal is to provide valuable insights while retaining accuracy. 
In particular, the formalism allows us to readily assess situations that often involve limited 
experimental resources: should one attempt a single high-precision measurement or would 
it be better to attempt two (or more) measurements with less precision\,\cite{Piekarewicz:2016vbn}. 

A particular interesting and topical example is the slope of the symmetry energy at saturation density 
$L$, a quantity that is closely connected to the pressure of pure neutron matter at saturation density. 
The slope of the symmetry energy plays an important role in areas as diverse as nuclear structure, 
heavy-ion collision, neutron star structure, and supernova 
explosion\,\cite{Tsang:2012se,Horowitz:2014bja,Thiel:2019tkm}. Given that $L$ is not a physical
observable, a determination of $L$ requires theoretical modeling. Yet, such a determination is 
hindered by the fact that different theoretical models tend to predict widely different values. 
However, the neutron skin thickness of ${}^{208}$Pb---defined as the difference 
in the root mean square radius between its neutron and proton distributions---has been shown 
to be strongly correlated to $L$\,\cite{Brown:2000,Furnstahl:2001un,Centelles:2008vu,
RocaMaza:2011pm}. The pioneering Lead Radius EXperiment (PREX) at the Jefferson 
Laboratory (JLab) has provided the first model-independent evidence in favor of a neutron-rich skin 
in ${}^{208}$Pb\cite{Abrahamyan:2012gp,Horowitz:2012tj}. Moreover, the recently completed 
PREX-II campaign aims to improve on the original PREX result by about a factor of three, thereby 
reaching a precision of about $0.06$\,fm on the neutron radius of ${}^{208}$Pb ($R_{n}^{208}$). 
Scheduled to run immediately after PREX-II, the Calcium Radius EXperiment (CREX) aims to 
determine the neutron radius of ${}^{48}$Ca ($R_{n}^{48}$) with a $0.03$\,fm 
precision\,\cite{CREX:2013}. Finally, the planned P2 experiment at the Mainz Energy recovery 
Superconducting Accelerator (MESA) will become the paradigm of a new generation of 
high-precision parity-violating electron-scattering experiments\,\cite{Becker:2018ggl}. Within 
the scope of the P2 setup, the Mainz Radius EXperiment (MREX) will determine 
$R_{n}^{208}$ and $R_{n}^{48}$ with a precision of $0.03$\,fm and $0.02$\,fm, 
respectively\,\cite{Thiel:2019tkm}. How will all these measurements impact the theoretical 
uncertainty in $L$ is a question that the formalism developed here is ideally positioned to 
answer.

\emph{Formalism.} Nuclear energy density functionals are calibrated by minimizing an objective 
``chi-square" function 
\begin{equation}
\ChiSq{N} ({\bf p}) = \sum_{n=1}^N \frac{\left(\On{n}{th}({\bf p}) - 
                                     \On{n}{exp}\right)^{2}}{\sigma_{n}^{2}},
\label{chi2}
\end{equation}
where $N$ is the number of experimental observables included in the fit, ${\bf p}$ is a point in the 
$F$-dimensional parameter space, $\On{n}{exp}$ represents the central value of the measured 
observables with an associated uncertainty $\sigma_{n}$, and $\On{n}{th}({\bf p})$ is the 
corresponding theoretical prediction. Although in principle $\sigma_{n}$ is associated with the 
experimental uncertainty, in practice it is often supplemented with a ``theoretical'' contribution. 
The main reason behind this choice is to prevent the model from being biased toward those 
observables that are measured with great precision, such as binding energies. 

The distribution of model parameters ${\bf p}$ informed by the $N$ experimental observables 
is encoded in the \emph{likelihood function} that represents the relative probability that a set 
of model parameters ${\bf p}$ reproduces the given experimental data. However, in general an 
arbitrary point in parameter space will often produce unphysical results or will lead to calculations 
that fail to converge. To overcome this situation one introduces a physically reasonable domain 
of parameters by incorporating one's own biases and intuition. 

Assuming that $\ChiSq{N} ({\bf p})$ is minimized at the point ${\bf p}_{{}_{0}}$, one can 
Taylor-expand Eq.(\ref{chi2}) around  ${\bf p}_{{}_{0}}$ to obtain
\begin{align}
  \ChiSq{N}({\bf p}) & \!=\! \ChiSq{N}({\bf p}_{{}_{0}}) \!+\!
  \delta p_i \!\left( \frac{\partial\ChiSq{N}}{\partial p_i}\right)_{\!\!0} \!+\!
  \frac{1}{2} \delta p_i \delta p_j \!\left( \frac{\partial^2\ChiSq{N}}
  {\partial p_i \partial p_j}\right)_{\!\!0} \!+\!\ldots \nonumber\\
  & \equiv
  \ChiSq{N}({\bf p}_{{}_{0}}) + \delta p_i {M}_{ij}\delta p_j +\ldots
 \label{chi2Taylor}
\end{align}
where $\delta p_i\!=\!({\bf p}\!-\!{\bf p}_{{}_{0}})_{i}$ and a summation over
repeated indices is assumed. The first term in the above expression 
is a constant while the second term vanishes at the minimum. Hence, the landscape near 
the minimum is entirely controlled by the curvature matrix $M$ (i.e., the matrix of second 
derivatives) which by construction is positive definite. In terms of the curvature matrix the 
covariance between two quantities $A$ and $B$ is given 
by\,\cite{Reinhard:2010wz,Piekarewicz:2014kza},
\begin{equation}
 \text{cov}(A,B) = \sum_{i,j=1}^{F} \left( \frac{\partial A}{\partial p_i} \right)_{\!\!0} 
 \text{\large $M$}^{-1}_{ij} \left( \frac{\partial B}{\partial p_j} \right)_{\!\!0},
 \label{covAB}
\end{equation}
where $M^{-1}$ is the covariance matrix. Given that the curvature matrix $M$ is 
positive definite, the existence of its inverse is guaranteed. If $A\!=\!B$, then the 
above expression yields the variance of $A$: $\text{\large{$\tau$}}_{\!\!A}^{2}\!=\!\text{cov}(A,A)$.
Note that we are using $\tau$ rather than $\sigma$ to denote the theoretical uncertainty;  
$\sigma$ is reserved to denote the experimental error. Finally, the Pearson correlation 
coefficient is given~by
\begin{equation}
 \text{\large$\varrho$}(A,B) = \frac{\text{cov}(A,B)}{\sqrt{\text{cov}(A,A)\cdot\text{cov}(B,B)}}\,.
 \label{corrAB}
\end{equation}
A value of $|\text{\large$\varrho$}(A,B)|\!=\!1$ implies that the two quantities are 
perfectly correlated. If instead $\text{\large$\varrho$}(A,B)\!=\!0$, then the two 
quantities are totally uncorrelated. In the particular case of the slope of symmetry energy 
$L$, we regard its central value and its associated theoretical uncertainty 
$\text{\large{$\tau$}}_{\text{\tiny$\!\!L$}}$ as our prior knowledge, given the $N$ 
experimental measurements. As new observables are incorporated into the data 
set, our knowledge of $L$ improves. 
 
\emph{Information content of one new observable}.
In this section we develop an analytic formalism to assess the information content of a 
new observable. The objective function resulting from the addition of one new observable 
is obtained by simply enlarging the sum displayed in Eq.(\ref{chi2}). 
There are (at least) two well-known approaches on how to estimate the impact 
of a new measurement. The simplest, yet numerically most costly, approach involves 
an entire new calibration of the model parameters\,\cite{Reinhard:2010wz}. A numerically 
less costly approach relies on Bayesian inference\,\cite{Gregory:2005,Stone:2013}. In this 
case one updates the distribution of model parameters using only the new available information. 
In the jargon of Bayesian inference, ``today's posterior is tomorrow's prior". Given that the prior 
distribution is exclusively a function of the model parameters, the likelihood function requires 
theoretical predictions for only the new observable. We introduce here a third approach that 
is entirely analytic. To do so, we now make the central assumption underlying our work: 
the minimum of the augmented objective function $\ChiSq{N\!\!+\!\!1}$ remains fixed at 
${\bf p}_{{}_{0}}$. Under this assumption, the second derivative of the augmented chi-square 
function becomes
\begin{equation}
 \left( \frac{\partial^{2}\ChiSq{N\!\!+\!\!1}}{\partial p_i\partial p_j}\right)_{\!\!0} \!=\!
 \left( \frac{\partial^{2}\ChiSq{N}}{\partial p_i\partial p_j}\right)_{\!\!0} \!+
 \frac{2}{\sigma_{\!I}^{2}}\!
  \left( \frac{\partial \On{I}{th}}{\partial p_i} \right)_{\!\!0}\!\!
  \left( \frac{\partial \On{I}{th}}{\partial p_j} \right)_{\!\!0},
  \label{SecondDeriv}
\end{equation}
where $\On{\!I}{}$ is the new experimental observable added to the previous data 
set. This relation implies that the augmented curvature matrix may be written as
\begin{equation}
 {\mathscr M}_{ij} = M_{ij} + \Gamma_{ij},
 \label{NewM}
\end{equation}
where the symmetric matrix $\Gamma$ has been defined as 
\begin{equation}
 \Gamma_{ij} = \text{\large$\gamma$}_{\!i}\text{\large$\gamma$}_{\!j} =
 \left[\frac{1}{\sigma_{{}_{\!I}}}\!\left(\frac{\partial \On{I}{th}}{\partial p_i} \right)_{\!\!0}\,\right]
 \left[\frac{1}{\sigma_{{}_{\!I}}}\!\left(\frac{\partial \On{I}{th}}{\partial p_j} \right)_{\!\!0}\,\right].
 \label{GMatrix}
\end{equation}
Given that the entire statistical framework relies on the inverse of the curvature matrix, we
must address how to invert the sum of two matrices, as given in Eq.(\ref{NewM}). In general,
the inverse of a sum of two matrices is not guaranteed to exist even when both matrices are
invertible. However, Miller has shown that in the case that one of the matrices has an inverse 
and the other one is a matrix of rank 1, then the inverse exists\,\cite{Miller:1981}. In our case
the augmented matrix ${\mathscr M}$ has been written as the sum of an invertible, positive
definite matrix $M$ plus a rank-1 matrix $\Gamma$ constructed from the outer product of 
the vector ${\bm \gamma}$ defined above. Note that the rank of a symmetric matrix is the 
dimension of its column (or row) space. The inverse of ${\mathscr M}$ is then given 
by\,\cite{Miller:1981}:
\begin{align}
 {\mathscr M}^{-1} &= M^{-1} - M^{-1} (g\Gamma) M^{-1}, \nonumber \\
 g^{-1} & = 1+{\rm Tr}(\Gamma M^{-1}).
 \label{Minverse}
\end{align}
The algorithm developed by Miller\,\cite{Miller:1981} for matrix inversion is analogous to 
the Sherman-Morrison-Woodbury formula that is better known to mathematicians and 
statisticians\,\cite{Banerjee:2014}. We adopt here Miller's approach as its generalization 
to the addition of an arbitrary number of observables is relatively straightforward.

Having obtained the augmented covariance matrix, we now compute the improved 
theoretical uncertainty in the slope of the symmetry energy $L$ (a quantity that we 
denote with a ``bar").  That is,
\begin{align}
 \text{\large{$\overline\tau$}}_{\!\!L}^{2} &= \sum_{i,j=1}^{F} 
 \left( \frac{\partial L}{\partial p_i} \right)_{\!\!0} 
 \text{\large${\mathscr M}$}^{-1}_{ij} \left( \frac{\partial L}{\partial p_j} \right)_{\!\!0} 
 \nonumber \\ & =
 \text{\large{$\tau$}}_{\!\!L}^{2} -g\sum_{i,j=1}^{F} 
 \left( \frac{\partial L}{\partial p_i} \right)_{\!\!0} 
 \Big(\text{\large${M}$}^{-1}\text{\large$\Gamma$}\text{\large${M}$}^{-1}\Big)_{ij}
 \left( \frac{\partial L}{\partial p_j} \right)_{\!\!0} \nonumber \\ & =
 \text{\large{$\tau$}}_{\!\!L}^{2}\left(1- g\frac{\text{\large{$\tau$}}_{\!\!I}^{2}}{\sigma_{\!I}^{2}} 
 \text{\large$\varrho$}^{2}(L,\On{I}{}) \right),
\label{TauBar0}
\end{align}
where $g$ is given by
\begin{equation}
 g^{-1} = 1+{\rm Tr}(\Gamma M^{-1}) = 1 + \frac{\text{\large{$\tau$}}_{\!\!I}^{2}}{\sigma_{\!I}^{2}}.
\end{equation}
In this way, one obtains the following simple and illuminating expression for the posterior variance of $L$:
\begin{equation}
 \frac{\text{\large{$\overline\tau$}}_{\!\!L}^{2}}{\text{\large{$\tau$}}_{\!\!L}^{2}}  =
 1 - \frac{\text{\large$\varrho$}^{2}(L,\On{\!I}{})}
 {1+\text{\large{$\sigma$}}_{\!I}^{2}/\text{\large{$\tau$}}_{\!\!I}^{2}} \equiv
  1 - \alpha_{I}^{2}\text{\large$\varrho$}^{2}(L,\On{\!I}{}).
\label{TauBar1}
\end{equation}
Note that the sole experimental contribution to this expression is $\sigma_{{}_{\!\!I}}$. Although many 
of the features encapsulated in the above expression are intuitive, the merit of the above expression
is that it is quantitatively precise. Indeed, once the central assumption underlying this work has been 
adopted, no additional assumptions or approximations are required to obtain Eq.(\ref{TauBar1}). In 
order to maximize the impact of the new measurement, the second term in Eq.(\ref{TauBar1})
should be made as large as possible. Evidently, if the new observable is perfectly correlated (or
anti-correlated) to $L$ and the experimental precision significantly improves on the current
theoretical uncertainty, then the reduction in the uncertainty is maximized. 

\emph{Information content of two new observables}.
Following the same exact procedure as before, one can write the augmented curvature matrix 
in the case that two new observables are added to the chi-square function. In this case the 
augmented curvature matrix is given by
\begin{equation}
 {\mathscr M}= M + \Gamma_{\!I} + \Gamma_{\!I\!I} \equiv {\mathscr M}_{I} + \Gamma_{\!I\!I},
 \label{NewM2}
\end{equation}
where $\Gamma_{\!I\!I}$ is the analog of Eq.(\ref{GMatrix}) for the second observable $\On{\!I\!I}{}$
and hence, a symmetric rank-1 matrix. Thus, the newly augmented matrix has the same structure as 
in Eq.(\ref{NewM}), namely, a positive definite invertible matrix ${\mathscr M}_{I}$ plus a rank-1 matrix 
$\Gamma_{\!I\!I}$. This guarantees that the new curvature matrix is also invertible: 
\begin{align}
  {\mathscr M}^{-1} & = {\mathscr M}_{I}^{-1} - 
    {\mathscr M}_{I}^{-1} (\widetilde{g}\,\Gamma_{\!I\!I}){\mathscr M}_{I}^{-1},  \nonumber \\
  \widetilde{g}^{-1} & = 1 + {\rm Tr}\!\left(\Gamma_{\!I\!I}{\mathscr M}_{I}^{-1}\right).
 \label{Minverse2}
\end{align}
Given that one is interested in assessing the impact of the two new measurements 
on prior theoretical uncertainties, one must compute the posterior covariance matrix 
${\mathscr M}^{-1}$, not in terms of ${\mathscr M}_{I}^{-1}$ but rather, in terms of $M^{-1}$. 
Because of space limitations, we only present here the main results and postpone a detailed
derivation to a forthcoming article. In terms of the original curvature matrix $M$ and the two 
additional rank-1 matrices $\Gamma_{\!I}$ and $\Gamma_{\!I\!I}$, the augmented covariance 
matrix is given by
\begin{widetext}
\begin{equation}
 {\mathscr M}^{-1} = M^{-1} - M^{-1} 
  \frac{\Big[\Big(\Omega_{I} + \Omega_{I\!I}\Big) - 
         \Big(\Omega_{I}M^{-1}\Omega_{I\!I} +\Omega_{I\!I}M^{-1}\Omega_{I}\Big)\Big]}
         {1-{\rm Tr}\Big(\Omega_{I}M^{-1}\Omega_{I\!I}M^{-1}\Big)} M^{-1},                        
\label{Minverse4}
\end{equation}
\end{widetext}
where we have defined 
\begin{equation}
 \Omega_{\lambda} \equiv g_{\lambda} \Gamma_{\!\lambda} = 
  \frac{\Gamma_{\!\lambda}}{1+{\rm Tr}\Big(\Gamma_{\!\lambda}M^{-1}\Big)};
 \quad (\lambda\!=\!I,I\!I).  
\label{BigOmega}
\end{equation}
Following the same steps as in Eq.(\ref{TauBar0}), one can now assess the impact of the
two new measurements on the theoretical uncertainty in $L$:
\begin{widetext}
\begin{equation}
 \frac{\text{\large{$\overline\tau$}}_{\!\!L}^{2}}{\text{\large{$\tau$}}_{\!\!L}^{2}} =
       1 - \left[
       \frac{\alpha_{I}^{2}\text{\large$\varrho$}^{2}(L,\On{\!I}{}) +
       \alpha_{I\!I}^{2}\text{\large$\varrho$}^{2}(L,\On{\!I\!I}{}) -
       2\alpha_{I}^{2}\alpha_{I\!I}^{2}\text{\large$\varrho$}(L,\On{\!I}{})
       \text{\large$\varrho$}(\On{\!I}{},\On{\!I\!I}{})\text{\large$\varrho$}(\On{\!I\!I}{},\!L)}
       {1-\alpha_{I}^{2}\alpha_{I\!I}^{2}\text{\large$\varrho$}^{2}(\On{\!I}{},\On{\!I\!I}{})}
        \right],
 \label{TauBar2}
\end{equation}
\end{widetext}
where $\alpha_{\!\lambda}^{2}$ has been defined in Eq.(\ref{TauBar1}). Besides the 
obvious generalization of the single term appearing in Eq.(\ref{TauBar1}) to two new
observables, there is an extra contribution from the correlation between the two new 
observables. In particular, if the two new observables are perfectly correlated, i.e., 
$\text{\large$\varrho$}(\On{\!I}{},\On{\!I\!I}{})\!=\!1$, then 
$\text{\large$\varrho$}(L,\On{\!I}{})\!=\!\text{\large$\varrho$}(L,\On{\!I\!I}{})$.
Further, if the ``first" observable is measured with high precision such that 
$\alpha_{I}\!\approx\!1$, then
\begin{equation}
 \frac{\text{\large{$\overline\tau$}}_{\!\!L}^{2}}{\text{\large{$\tau$}}_{\!\!L}^{2}} 
= 1- \text{\large$\varrho$}^{2}(L,\On{\!I}{}).
 \label{TauBar4}
\end{equation}
This expression is identical to Eq.(\ref{TauBar1}) in the appropriate limit, so the ``second" 
measurement becomes superfluous.

We have mentioned that the entire procedure hinges on the fact that the augmented curvature 
matrix may be written as the sum of an invertible curvature matrix plus a rank-1 matrix. This fact 
alone guarantees that the augmented curvature matrix is invertible. Thus, this recursive procedure 
can be generalized to an arbitrary number of new measurements\,\cite{Miller:1981}, a topic that 
will be addressed in a forthcoming publication.

\emph{Results.} Given that many critical quantities associated with the equation of state are not 
genuine physical observables, we now provide a couple of examples on how future experiments 
and observations may be used to improve the theoretical uncertainties in such quantities. 

By the end of 2020 analyses of the PREX-II and CREX campaigns are expected to be completed. 
The expectation is that these experiments will reach a precision of $0.06$\,fm for $R_{n}^{208}$
and of $0.03$\,fm for the corresponding radius in ${}^{48}$Ca. Adopting these two experimental 
errors as the current theoretical uncertainties, we want to examine the impact on $L$ from MREX 
at MESA, which aims to determine $R_{n}^{208}$ and $R_{n}^{48}$ with a precision of 
$0.03$\,fm and $0.02$\,fm, respectively\,\cite{Thiel:2019tkm}. 
To assess the impact of MREX one needs a model to connect the two physical observables 
($R_{n}^{208}$ and $R_{n}^{48}$) to the unobservable slope of the symmetry energy. 
To do so, we employ the covariant energy density functional FSUGold2\,\cite{Chen:2014sca}
that was calibrated using exclusively well-measured observables, namely, ground-state binding 
energies and charge radii, centroid energies of giant monopole resonances, and observational 
limits on the maximum neutron star  mass. FSUGold2 predicts the following correlation coefficients 
for the quantities of interest: $\varrho(R_{n}^{208},R_{n}^{48})\!=\!0.963$, 
$\varrho(L,R_{n}^{208})\!=\!0.966$, and $\varrho(L,R_{n}^{48})\!=\!0.921$.
Using these predictions, one can now estimate the reduction in the theoretical uncertainty in $L$ from 
improved measurements of neutron radii. 
\begin{figure}[ht]
 \centering
 \includegraphics[width=0.4\textwidth]{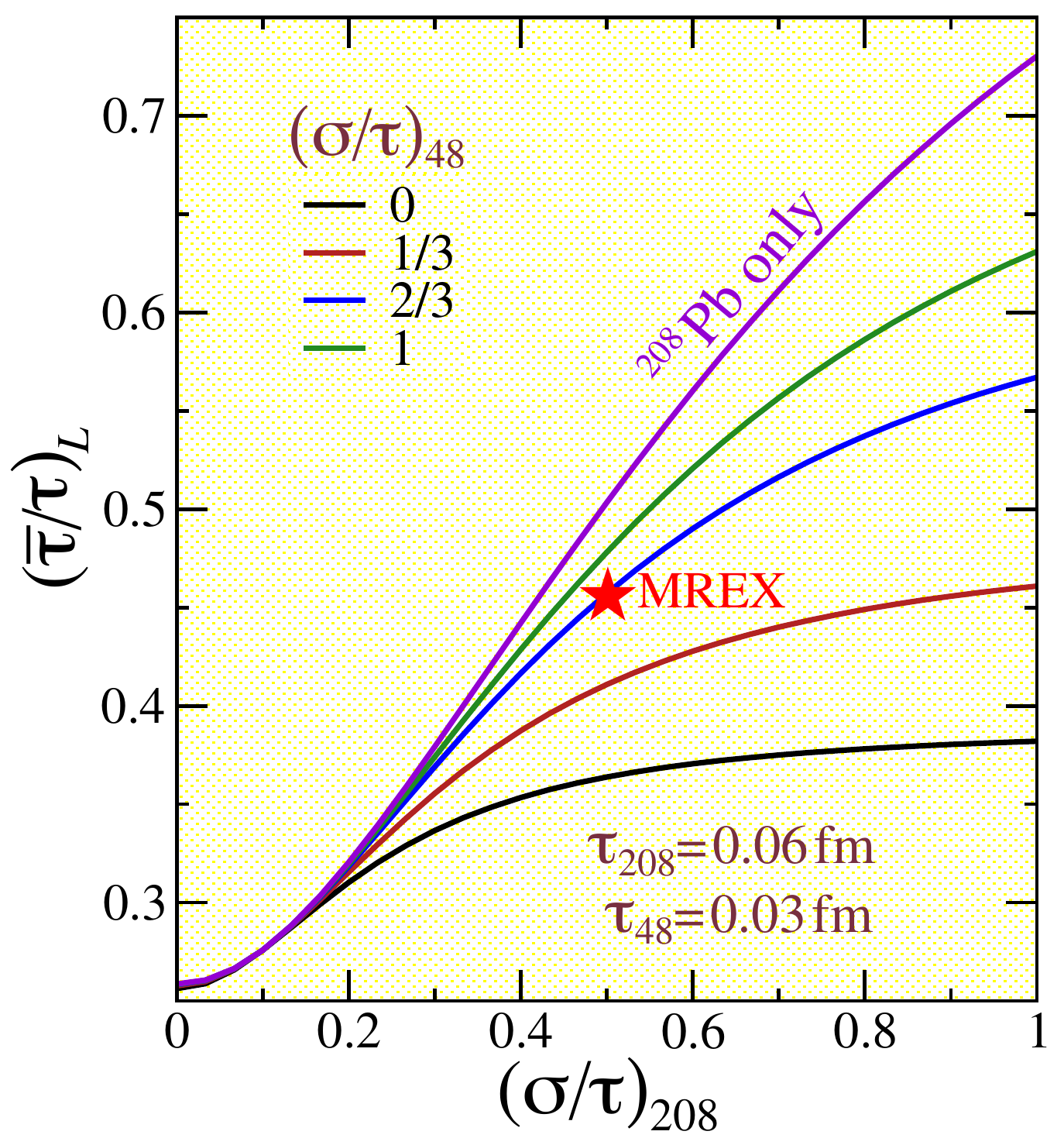}
 \caption{(Color online) Reduction in the theoretical uncertainty of the slope of the symmetry 
  energy $L$ as a consequence of improved measurements of the neutron radius of both 
  ${}^{208}$Pb and ${}^{48}$Ca. The star in the figure reflects the impact of MREX on the slope 
  of the symmetry energy.}
\label{Fig1}
\end{figure}
Results are displayed in Fig.\ref{Fig1} as a function of the 
fractional experimental error ($\sigma_{208}/\tau_{208})$ for a few values of the corresponding 
quantity in ${}^{48}$Ca. The ``star" in the middle indicates
the more than a factor of two improvement in the theoretical uncertainty in $L$ under the assumption that
MREX reaches its precision goals. Also shown in the figure is the result obtained from assuming that only  
$R_{n}^{208}$ is measured. The figure suggests that one could achieve the projected MREX theoretical 
uncertainty in $L$ from a slightly improved single measurement of $R_{n}^{208}$. We underscore that 
although the numbers adopted for this example are representative of our current understanding, the 
main reason behind this---and the next example---is to illustrate the insights encapsulated in 
Eq.(\ref{TauBar2}). 

In the next example we illustrate the impact of new experiments and observations on another 
quantity that, while critical to our understanding of neutron-rich matter, is not accessible to
experiment: the pressure of pure neutron matter at two times saturation density $P_{2}$. 
Twice nuclear matter saturation density provides a powerful bridge between terrestrial 
experiments and cosmological observations. At terrestrial facilities neutron-rich matter at 
above saturation density can be probed via energetic collisions of heavy ions with a large 
neutron-proton asymmetry\,\cite{Tsang:2008fd,Horowitz:2014bja,Thiel:2019tkm,
Tsang:2019mlz,FRIB400:2019}. In the cosmos, the pressure of neutron-rich matter at 
supra saturation density can be elucidated from the structure of neutron stars. Recent 
detections by both the LIGO-Virgo and NICER collaborations have started to provide constraints 
on the compactness of neutron stars and ultimately on stellar radii\,\cite{Bauswein:2017vtn,
Fattoyev:2017jql,Annala:2017llu,Abbott:2018exr,Most:2018hfd,Tews:2018chv,Malik:2018zcf,
Tsang:2018kqj,Radice:2017lry,Radice:2018ozg,Xie:2020tdo}. In turn, stellar radii seem to 
be strongly correlated to the pressure of pure neutron matter at two-to-three times saturation 
density\,\cite{Lattimer:2006xb,Tews:2019cap,Capano:2019eae,Drischler:2020yad}. 
For a comprehensive recent review on equation of state constraints see 
Ref.\,\cite{Chatziioannou:2020pqz} and references contained therein.

\begin{figure}[ht]
 \centering
 \includegraphics[width=0.4\textwidth]{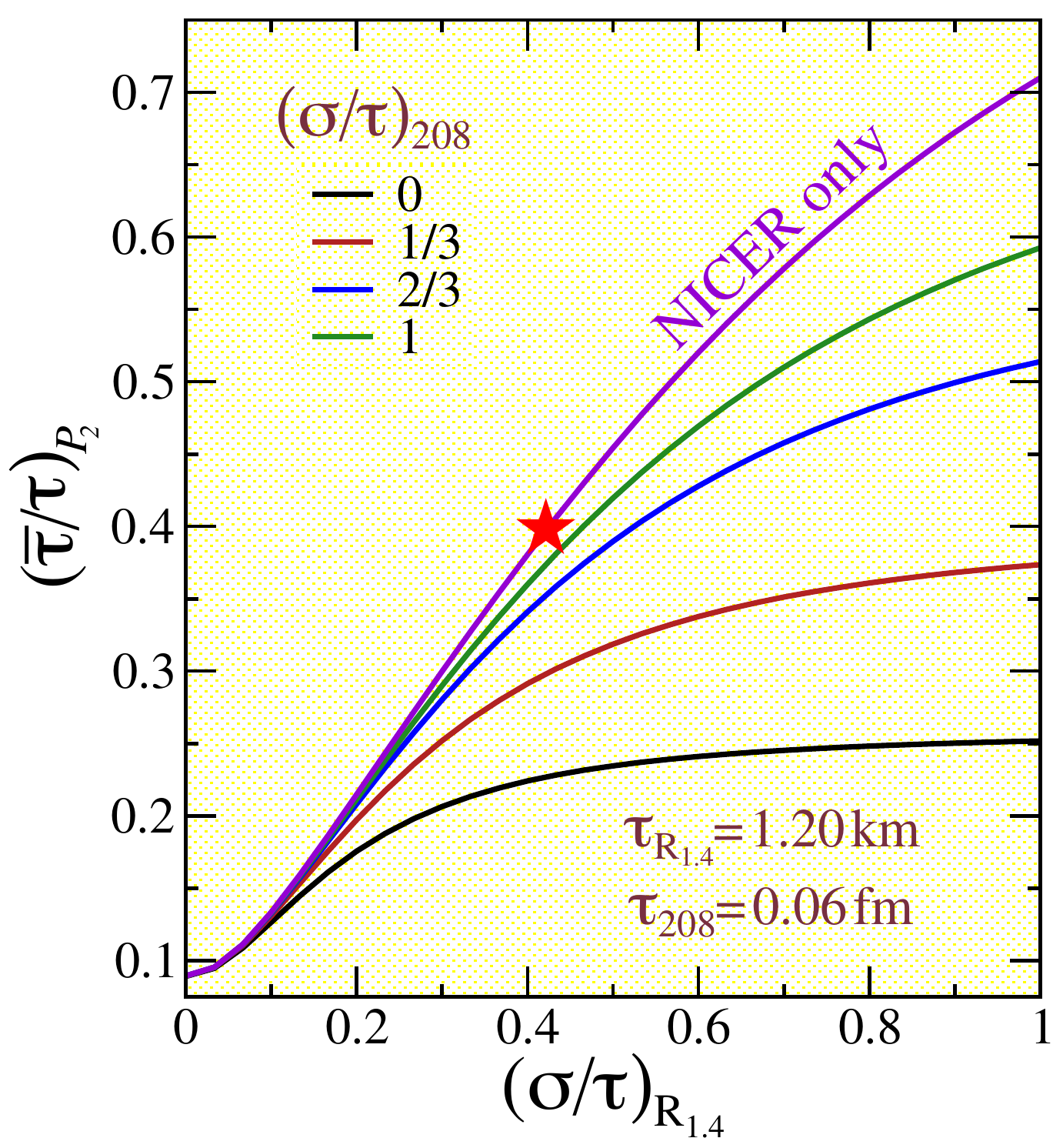}
 \caption{(Color online) Reduction in the theoretical uncertainty of the pressure of pure neutron
  matter at twice nuclear matter saturation density as a consequence of improved measurements 
  of the radius of a $1.4\,M_\odot$ neutron star and the neutron radius of ${}^{208}$Pb. The star 
  in the figure reflects the impact of a 0.5\,km measurement of the radius of J0437-4715.}
\label{Fig2}
\end{figure}

In Fig.\ref{Fig2} we display the reduction in the theoretical uncertainty in the pressure of
pure neutron matter at twice saturation density ($P_{2}$) as a consequence of improved 
measurements of stellar radii and neutron skins. The theoretical uncertainty in the radius 
of a $1.4\,M_\odot$ neutron star is set to $1.2$\,km\,\cite{Miller:2019cac,Riley:2019yda}. 
As done earlier, we used predictions from FSUGold2 for the correlation coefficients of 
interest: 
$\varrho(R_{1.4},R_{n}^{208})\!=\!0.968$, 
$\varrho(P_{2},R_{1.4})\!=\!0.996$, and 
$ \varrho(P_{2},R_{n}^{208})\!=\!0.966$.
Figure \ref{Fig2} shares many similar features to Fig.\ref{Fig1}. However, the slightly larger correlation 
coefficients suggest that a ``perfect" measurement of the stellar radius could reduce the theoretical 
uncertainty in $P_{2},$ by nearly 90\%. The star in the figure represents the impact of a 
$\pm\,0.5\,{\rm km}$ determination of the stellar radius of J0437-471\,\cite{Miller:2019cac,Riley:2019yda},
a neutron star with a well determined mass of 1.44(7)\,$M_{\odot}$\,\cite{Reardon:2015kba}. By itself, 
this measurement would reduce the theoretical uncertainty in $P_{2}$ by about 60\%.

We close this section with a word of caution. Whereas the strong correlation between the neutron skin 
thickness of ${}^{208}$Pb and $L$ has been firmly established\,\cite{RocaMaza:2011pm}, we find that 
the correlation between the radius of a $1.4\,M_\odot$ neutron star and the pressure of pure neutron 
matter at twice saturation density is model dependent. Indeed, accurately calibrated models with a soft 
symmetry energy, such as FSUGarnet\,\cite{Chen:2014mza}, predict a correlation coefficient of only
$\varrho(P_{2},R_{1.4})\!=\!0.4$, even if the entire equation of state---containing contributions from 
both symmetric nuclear matter and the symmetry energy---is stiff enough to support two solar-mass 
neutron stars.

\emph{Conclusions.} Motivated by one of the central questions posed in Ref.\cite{Reinhard:2010wz},
namely, ``How can new data reduce uncertainties of current theoretical models?", we developed a
fully analytic approach to answer this question. Our entire formalism hinges on one underlying 
assumption, namely, that the location of the minimum of the chi-square function remains unchanged 
after incorporating the new set of observables. Once this assumption is adopted, simple and 
insightful expressions were developed as one or two observables are added to the calibration of the 
model parameters. Precise details on the derivation of our main results, encapsulated in 
Eqs.(\ref{TauBar1}) and (\ref{TauBar2}), and their generalization to the case of an arbitrary number 
of new observables will be presented in a longer forthcoming publication.

Given the relative simplicity of Eqs.(\ref{TauBar1}) and (\ref{TauBar2}) we suspect that such expressions 
are well known to mathematicians and statisticians. Yet despite our best efforts---and that of several
colleagues---we could not find such a result in the literature. Although we found countless 
applications of the Sherman-Morrison-Woodbury formula, we found no answer to the question of
how can new data reduce uncertainties of current theoretical models.

\emph{Acknowledgments.}
 We thank Pablo Giuliani for a careful reading of the manuscript and for many useful comments.
 We also thank Jonathan Bradley and Antonio Linero for many useful conversations. This 
 material is based upon work supported by the U.S. Department of Energy Office of Science, 
 Office of Nuclear Physics under Award Number DE-FG02-92ER40750 .
\vfill\eject

\bibliography{PRCRapid.bbl}
\end{document}